\documentclass[a4paper,11pt]{article}
\usepackage{pos}

\title{Two Decades of Education and Public Outreach with Chicago Public Schools}
 \ShortTitle{Education and Public Outreach with Chicago Public Schools}
\author*[a]{Vikram V Dwarkadas}

\affiliation[a]{Department of Astronomy and Astrophysics, University of Chicago,\\
  5640 S Ellis Ave, Chicago, IL, 60637}

\emailAdd{vikram@astro.uchicago.edu}
\abstract{Over the past two decades, I have been actively involved in teaching astronomy and astrophysics to Chicago Public School (CPS) students and their teachers, in collaboration with various groups as well as by myself. Valuable resources that we have created for schools include the Multiwavelength Astronomy Website, with modules for infrared, optical, ultraviolet, X-ray and gamma-ray astronomy. The content of each lesson is derived from interviews with scientists, archived oral histories, and/or memoirs. Lessons are evaluated by a science educator and at least one subject matter expert before being produced for the web. They are supplemented by NASA media, archival material from the University of Chicago Library and other archives, and participant contributed photographs, light curves, and spectra. Summer programs provided training to CPS teachers to use the resources in their classrooms.  Currently, I lead the Chicago Area Research Mentoring (CHARM) initiative. In the past academic year I worked with a class of 17 diverse 11th grade honors students at the University of Chicago Charter School, Woodlawn. Through frequent lectures ($\sim$ every 4 weeks), these students were exposed to astrophysical topics and concepts not normally covered in a school curriculum. CHARM aims to develop the student's critical thinking, introduce them to astrophysical research methods and techniques, and prepare them for a career in science, technology, engineering and mathematics (STEM), particularly a research-oriented one. In this article, I highlight some projects, educational resources, results achieved, and lessons learned along the way.}

\FullConference{37$^{\rm{th}}$ International Cosmic Ray Conference (ICRC 2021)\\
		July 12th -- 23rd, 2021\\
		Online -- Berlin, Germany}

\def\gtrsim{\mathrel{\hbox{\rlap{\hbox{\lower4pt\hbox{$\sim$}}}\hbox{$>$}}}}

\newcommand\be {\begin{equation}}
\newcommand\en{\end{equation}}

\newcommand{\ee}{\end{equation}}

\newcommand{\beq}{\begin{equation}}
\newcommand{\eeq}{\end{equation}}
\newcommand{\bea}{\begin{eqnarray}}
\newcommand{\eea}{\end{eqnarray}}
\newcommand{\bei}{\begin{itemize}}
\newcommand{\eei}{\end{itemize}}
\newcommand{\bee}{\begin{enumerate}}
\newcommand{\eee}{\end{enumerate}}

\begin{document}
\maketitle

In order to increase science, technology, engineering and mathematics (STEM) literacy, it is important that students start learning science, and begin to appreciate science and the scientific method, from a young age. This requires that science, the scientific method, and the essential mathematics, be taught in middle and high schools, at an age-appropriate level.  To achieve this goal, for almost two decades I have worked on education and public outreach projects at the University of Chicago. I have collaborated with several groups, especially with Prof.~Don York (UChicago) to create resources for Chicago Public Schools (CPS) from an astronomical perspective, train CPS teachers to use the material, and deploy it effectively in their classrooms.  

In the following, I outline some of the projects that I have been, and continue to be, involved in. I give an overview of the projects, what was achieved and, where possible, what participants thought.  I point to some freely available resources resulting from these that are useful to everyone, not just the demographic that they were designed for. At the end, I describe some lessons learned, and offer suggestions for what can be done better. 

It is important to point out here that every project that I have been a part of had to be independently funded. We submitted grant proposals to funding agencies for our projects. If our proposal was accepted and funded adequately, we were able to successfully carry out our projects. It is certainly easier if you are part of a larger organization supporting the outreach, and providing a constant source of funding that helps to maintain staff. This was not the case for the projects that I have participated in.\\

{\bf Teaching Resources for Chicago Public Schools:} With funding from Education and Public Outreach (E/PO) grants from various sources, we have created  interactive resources for students in CPS schools, but available on the internet for anyone to use.

\begin{itemize}
\item Under the auspices of a Space Telescope Science Institute Education and Public Outreach grant in 2006, my collaborators and I created a website for students that provides a clear understanding of the formation of the elements in the periodic table. We outlined the animations and simulations that would work with the scientific concepts, and wrote the appropriate text to accompany the graphics. Our graphics designer then rendered the animations to accompany the text.  It was designed using Adobe Flash, as was the norm in those days, and therefore may not work unless you have older computers with web browsers running Flash. The interactive (Figure \ref{fig:oote}) is available at the following website: \\
{\it http://new.web-docent.org/modules/interactives/misc/originsoftheelements.swf}

%second sentence below is confusing.
\begin{figure}[htbp]
\includegraphics[width=1.\columnwidth]{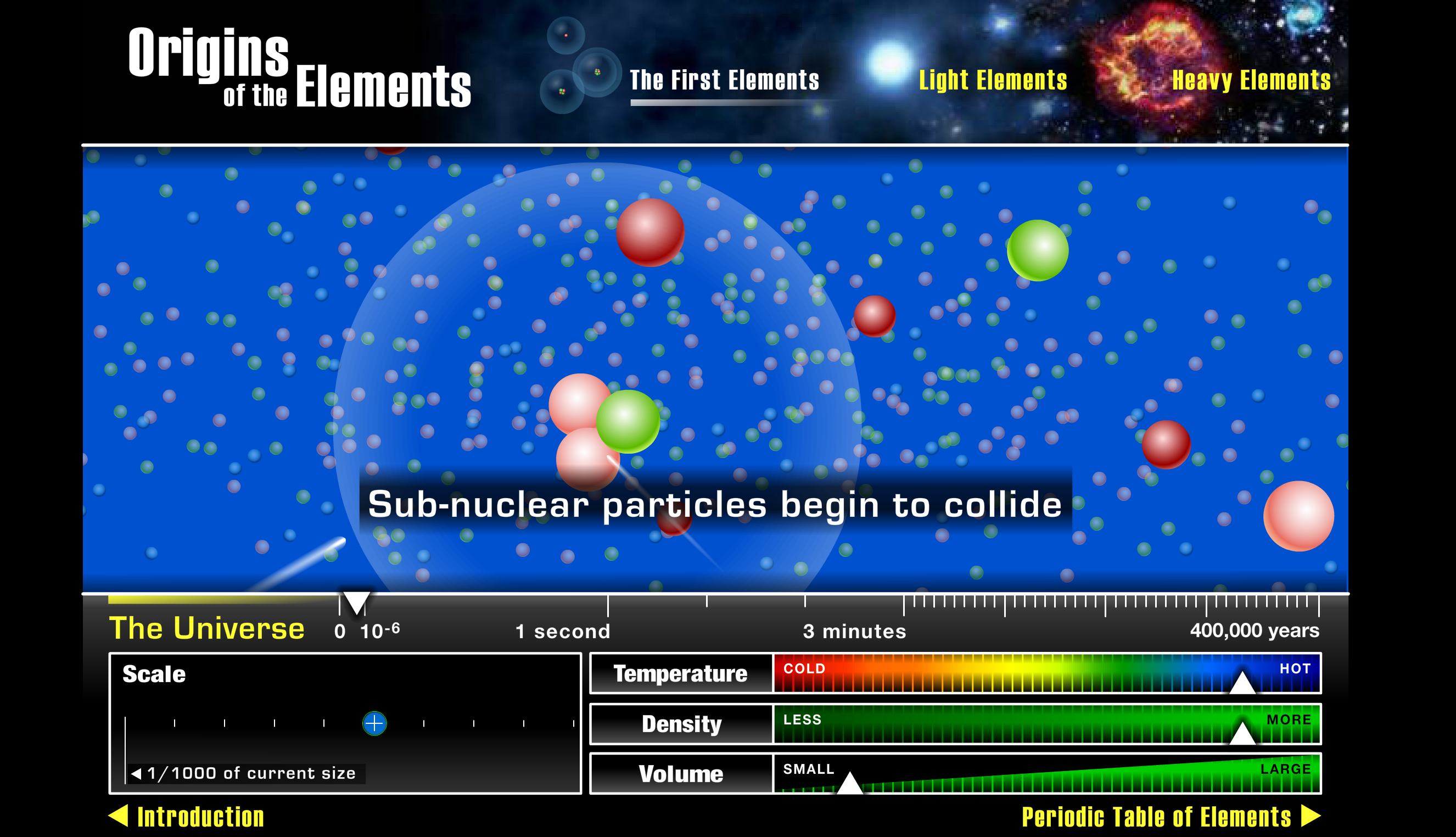}
\caption{Origin of the Elements website. This was created in Adobe Flash, and may no longer runs unless you still have an older computer running Flash.
\label{fig:oote}}
\end{figure}

\item We followed this up with a series of NASA grants focussed on the development of a valuable and instructive teaching resource on {\bf multiwavelength astronomy}. The project lasted the better part of a decade. It was geared towards developing modules for the various wavelength regimes comprising the electromagnetic spectrum, training teachers to use the resource, and deploying it in Chicago public schools. The first grant supported optical, infra-red and ultraviolet astronomy, while the next was aimed at X-ray and gamma-ray astronomy\footnote{We had planned on a future proposal to include radio astronomy and gravitational waves, but the NASA E/PO Program was closed before that could be accomplished.}. We finished with the NASA All-Stars program aimed at developing teachers' in-depth content area knowledge, pedagogical training for science inquiry, providing state-of-the-art learning opportunities to do astronomical inquiry, and ongoing mentoring. I was Co-I on the X-ray/Gamma-ray grant, and a subject matter expert on the NASA All-Stars program.

One outcome of  our work was the "Multiwavelength Astronomy" website.  The content of each lesson was derived from interviews with scientists, archived oral histories, and/or memoirs. Lessons were evaluated by a science educator and at least one external subject matter expert before being produced for the web (Figure \ref{fig:mwastro}). They were supplemented by NASA media, archival material from the University of Chicago Library and other archives, and participant contributed photographs, light curves, and spectra. 

\begin{figure}[htbp]
\includegraphics[width=1.\columnwidth]{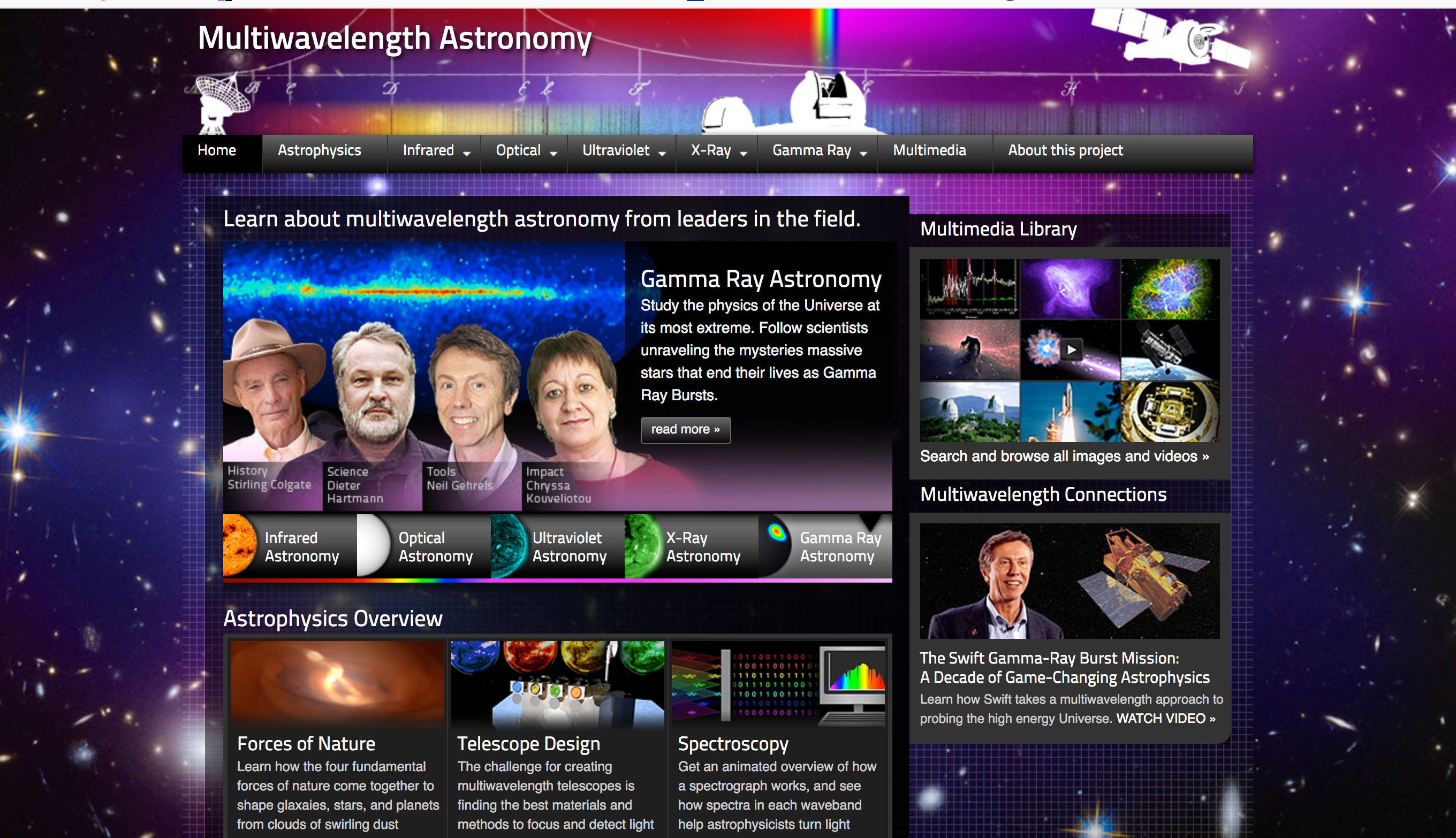}
\caption{The home page of the multiwavelength astronomy website.
\label{fig:mwastro}}
\end{figure}

% Our grant requests included funding for a computational device that they could use during the program, and they were allowed to keep at the end of the program.

Accompanying the development of the lessons, a 2--3 week summer program was held, including CPS teachers and some of their students (Figure \ref{fig:nasastars}). During these programs, the teachers and students  were introduced to the lessons, and trained in astronomy at various wavelengths.  Students were invited to participate so that the effectiveness of the lessons could be gauged in practice. The students came from different backgrounds, and were not all expected to have the necessary computational background, and/or equipment, to tackle the scientific projects that were taught. Our grant included funding for a computational device that they could use during the program, and they were allowed to keep after the program ended. One year, for example, this was an Ipad Mini, another year, it was a 10-inch ASUS laptop with a 64 GB flash drive running a full version of Windows. Projects and instruction were geared to the particular device provided so that students would learn to make the most use of it, and be able to use it in their own schoolwork. The lessons created were taught by the CPS teachers with our help, and students would give valuable feedback on how they could be improved. The students also carried out research under their teachers and our supervision, and presented their projects at the end. 

\begin{figure}[htbp]
\includegraphics[width=1.\columnwidth]{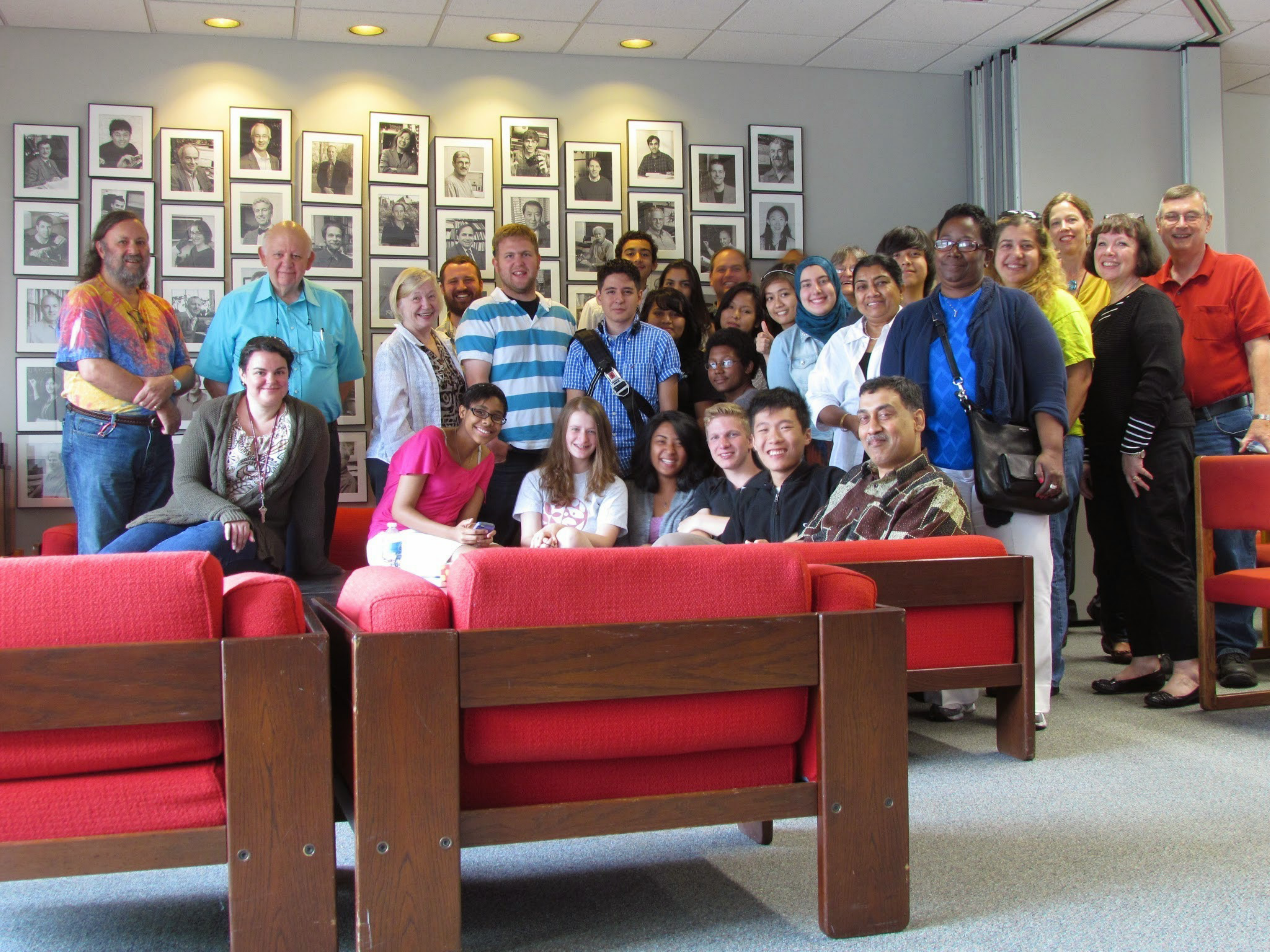}
\caption{A picture (taken by me) showing the group at the NASA  summer program of July 2013. This image shows all organizers, the CPS teachers, students, and the course evaluators. That's me in the front in the brown shirt. I literally had to jump over the couch after setting the 10-second delay timer to take the photograph (but this ensured that everyone is smiling :))
\label{fig:nasastars}}
\end{figure}

Participants at the ICRC may be particularly interested in the Gamma-ray section. The history of the field is covered by Stirling Colgate. Stirling had interesting stories to tell, not only involving supernovae and gamma-rays but also how he contributed to peace efforts with Russia. These are the kind of stories that students are always inspired to hear. An introduction to gamma-ray science is wonderfully presented by Dieter Hartmann. The development of tools to understand gamma-rays, and how the field advanced, is expertly written by Neil Gehrels. Finally, the impact of gamma-ray astronomy till that time is brilliantly outlined by Chryssa Kouveliotou. This section emphasizes gamma-ray bursts and related phenomena, which were being heavily studied when the content was being created. The lives of the various astrophysicists, and the different paths that they took to reach their current status, were all appreciated by the students as much as the science that they were introduced to.

A few practical considerations: The funding request needed to include a stipend for the teachers to facilitate their 2--3 week attendance, as well as all staff necessary to man the event. Lunch for the students and teachers was also provided. If the students were at school, lunch would have been similarly provided. The classrooms did not necessarily have a charging station available for each student. In order to ensure that the devices were fully charged at the beginning of the day to enable day-long use, the staff would collect the devices every day, charge them overnight and return them to the students at the beginning of the next day. Keeping them on site until the end of the program also provided incentive to the students to attend every day. 

Students met and interacted with some of the experts responsible for the lessons. According to their blogs, they really enjoyed this aspect, since they got to interact with leaders in the field and personally ask them questions. The lessons themselves need to be at an appropriate level. The students would comment on the day's lessons in their blogs, so if something was not up to standard, it was quickly brought to our attention. We suggest that writing of daily blog posts should  be encouraged, and an appropriate blogging medium set up.

The interactive multi-wavelength astronomy site can be accessed at: \\ {\it http://ecuip.lib.uchicago.edu/multiwavelength-astronomy/}

I personally gave talks to CPS teachers and students on cosmic explosions, as well as X-ray astronomy. In July 2010, my 1-hour talk to CPS teachers turned into a 4-hour marathon instruction session on X-ray astronomy. In July 2013, another 1.5 hour talk to CPS teachers and their students was labelled by one of the students as his ``favorite lecture." Another lecture on cosmic explosions was delivered on July 2014 to CPS teachers and students. It was clear from the experience that while the students are not exposed to such topics in their curriculum, they are certainly eager to learn. 

The immediate impact of the program can be gleaned from some of the online blog posts. It was clear that many students had learned more from the program than they had imagined at the start. On the last day, several bemoaned the fact that the program was ending. It seemed to have inspired many of them to study science, which was the goal. For us, a source of pride is that one of the students from the programs is now a graduate student in our department of Astronomy and Astrophysics at the University of Chicago.
 
\begin{figure}[htbp]
\includegraphics[width=1.\columnwidth]{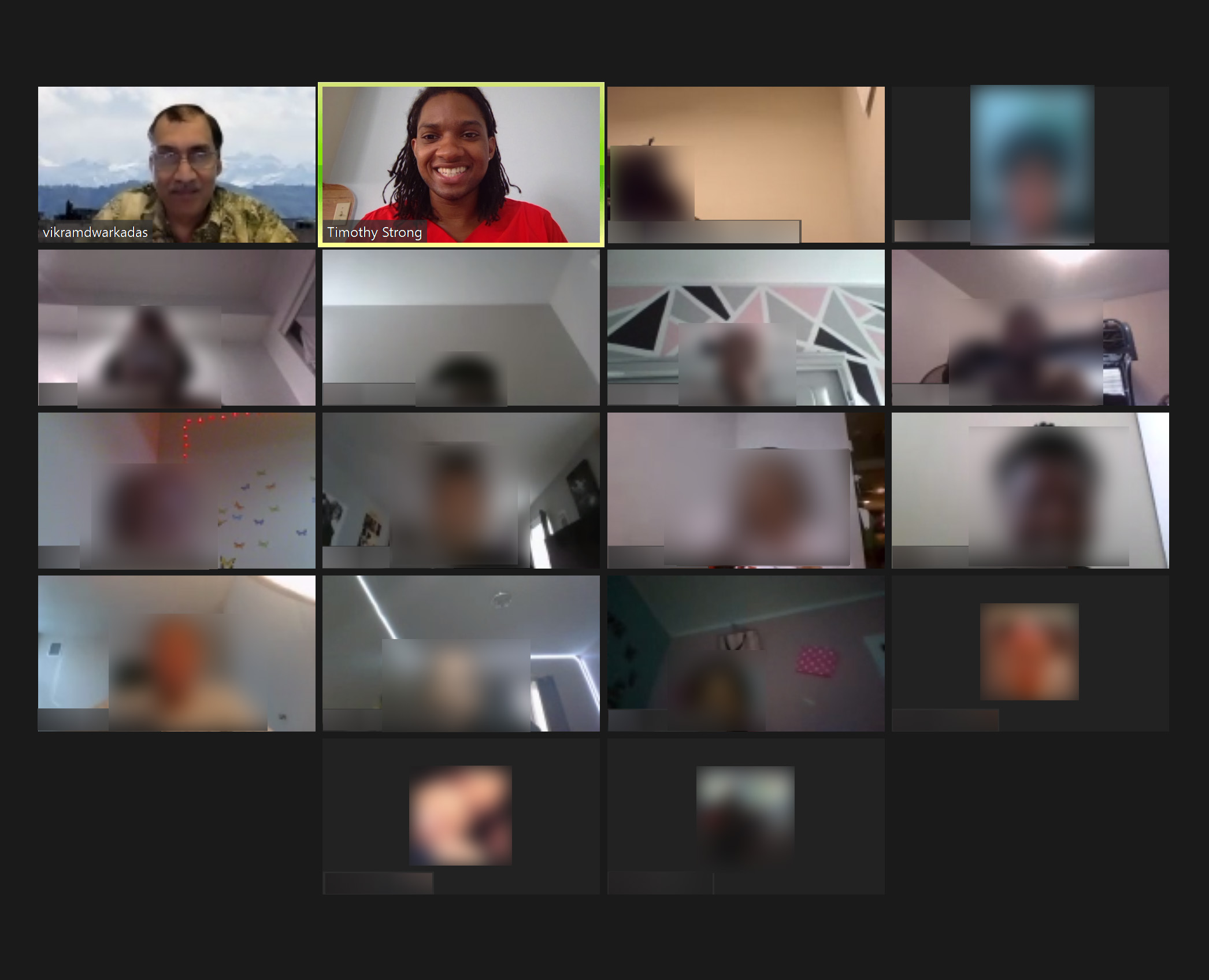}
\caption{A (zoom) group photo, taken by the class teacher Mr.~Strong, showing him, some of his students, and me, in the grade 11 Physics honors  class that I was teaching during the past academic year at the UC Woodlawn school, Chicago. Reproduced with permission granted by Mr.~Strong. Images of students blurred on arxiv.
\label{fig:charm}}
\end{figure}

{\bf Chicago Area Research Mentoring (CHARM) project:} In the past academic year I taught astronomy ($\sim$ every 4 weeks) to 17 students in the 11th grade Physics Honors class at the UC Woodlawn charter school, part of the Chicago Public Schools
system (Figure \ref{fig:charm}).  I taught topics in stellar evolution, the deaths of stars, especially massive stars, and supernovae, 
which students are generally not exposed to. The aim was to develop critical thinking skills, and introduce them to
research methods and techniques. Care must be taken to give talks at an age-appropriate level, while still covering the basic science. That's a difficult line to tread, but I have found that giving analogies to everyday phenomena really helps. For example, I have talked about hydrodynamic instabilities such as Rayleigh-Taylor and Kelvin-Helmholtz with high school students, not a topic that they would normally hear about. This comes up for instance when they see structure and filamentation in cosmic phenomena. Although this is initially confusing, when it is related to earthly phenomena such as stalactites, water dripping from the ceiling, cream over coffee, and jet trails,  the students begin to see the physics in action, and begin to realise that the the physics of the cosmos is not necessarily all that different from that which shapes phenomena on earth. 
 \end{itemize}

\underline {\bf Discussion and Lessons Learned} 
\label{S:DIS}

\begin{itemize}
\item It is important to educate students in STEM topics from a  young age. It is  necessary to have age-appropriate teaching, while recognizing that the backgrounds and skills of the students are quite different.  Many of these students are extremely capable, read a lot, and are aware of important discoveries in the news. It is our job to nurture and encourage their passion.
\item Multiwavelength astronomy is not generally taught at the high school and middle school level, but there's no reason why it cannot be. At most, students may learn about optical astronomy, but almost never venture into high energy astronomy including X-ray and gamma-ray wavelengths. This is specifically why we created the multiwavelength astronomy website. It is at an appropriate level for middle and high schoolers. Please use it in your own teaching.
\item In order to encourage a diverse and growing STEM population, governments, and perhaps private organizations, need to find a way to support it while ensuring equitable access.  Many students in our summer programs could not have participated without the available funding, and the computational device that was handed to them, which they used for the majority of their work. The NASA funding allowed students of all backgrounds to participate equally.
\item In this context, it is crucial that funding agencies  provide funds not only to create educational resources, but to maintain and update them. Programming languages, platforms, and display formats will change over time. Educational interactives and animations created with Adobe Flash over the years are now unusable. Updating still pertinent resources to harness the latest technologies should be considered essential. While grants are rightfully awarded for innovative new projects, the maintenance of relevant existing programs, and the updating of resources to incorporate newer technologies should also be given consideration for funding. 
\item Scientists need to find the time to personally talk to, teach, and encourage young minds. The picture that some students have is of scientists being smart but aloof and unapproachable, a notion that we need to dispel.  {\em The student-teacher relationship is clear, but that distinction does not mean playing down their questions. Instead we need to allow for their curiosity while answering their questions. Students can quickly tell who they would like to work with and learn from, and when their opinions are valued. A good working relationship leads to them being more invested in their work.} While this is true at all stages, it is especially true for younger students. 
\item It is important to spend time discussing with students not only about getting into science programs, but what it entails, what the statistics are and what the prospects are. There is no need to sugarcoat difficult paths. Students like to know for themselves how easy or difficult it is for them to advance in the field, and what they are getting themselves into. At the same time, they need to be made aware of alternatives should their chosen path not work out, and be secure in the knowledge that they can use their training, and especially the problem-solving skills that they have acquired, in many different fields.
\item Do not avoid mathematics. Explain it so that students can actually understand the reason for an equation and why every term is there. Students do not start out knowing that mathematics is difficult, but they are sometimes told to think that way. Once they realise that with simple mathematics they can understand every day phenomena, they begin to like it.
\item Describing the history of the topic and interesting aspects of the life of the individuals involved is important. Students need to grasp that the individuals who made some of the most important discoveries were once students like them, they were not born with the knowledge.
\item It is important to highlight the process of scientific discovery and the amount of time discoveries take. Students nowadays see wonderful pictures taken with space satellites. After I gave a talk on X-ray astronomy, its history and various X-ray satellites, when students were asked what they learned, many of them seemed awed by the fact that a satellite like Chandra essentially took almost 30 years from definition of the concept till it was launched. They were amazed that one person, or one team, would devote so much time to a single project, with no guarantee of success. Considering their young age, the fact that some people had spent more than their lifetime working on a single project was astonishing and difficult to easily comprehend, and viewed with a sense of wonder. Major telescopes take decades to design and build. Providing students with those time-lines makes them appreciate the effort involved, provides a sense of scale, and puts the images into perspective. 
\item It is important in teaching students to make them think on their own, via frequent questions, and develop the critical thinking skills necessary to tackle tough problems in any field. 
\item The younger generation loves to blog. Encourage blogging. Many are more apt to write about their thoughts and feelings freely from behind a computer screen, than actually state them aloud in class. It is a good way to get feedback and know what they are thinking.
\item The younger generation has grown up using technology, and loves to embrace it.  It was amazing to see how quickly they adapted to a new Ipad, or any new device. They soon learned how to use it for various purposes, even teaching their teachers a few tricks. And yes, sometimes it is likely that they will use it for fun. As long as they use it also for work, that should be fine. Our summer programs were designed to make learning fun.

\end{itemize}

{\bf Acknowledgements:} VVD's work is currently supported by NSF Grant 1911061. The NASA summer programs and multiwavelength astronomy website were funded by NASA EPO Grants NNX09AD33G and  NNX10AE80G. I would like to thank Don York, from whom I have learned a lot over the years. This work would have not been possible without some very helpful and dedicated staff members, including Julia Brazas, Christie Thomas, Shaz Rasul and Sara Dennison.

\bibliographystyle{JHEP}
\bibliography{my-bib-database}

%% Full authors list (ONLY FOR COLLABORATIONS)
%\clearpage
%\section*{Full Authors List: \Coll\ Collaboration}
%
%\noindent \textbf{Note comment afterwards:} Collaborations have the possibility to provide an authors list in xml format which will be used while generating the DOI entries making the full authors list searchable in databases like Inspire HEP. For instructions please go to icrc2021.desy.de/proceedings or contact us under icrc2021proc@desy.de.\\
%
%\scriptsize
%\noindent
%first.author$^1$, 
%second.author$^2$, 
%third.author$^3$ % .... more names
%and 
%last.author$^{n}$ \\
%
%\noindent
%$^1$first.affiliation.
%$^2$second.affiliation. % .... more affiliation
%$^{m}$last.affiliation.

\end{document}